\documentstyle[aps,preprint]{revtex}
\input{epsf}
\begin{document}
\title{Ab initio calculations of the hydrogen bond}
\author{ B. Barbiellini$^{1}$,
A. Shukla$^{2}$}
\address{$^{1}$Dept. of Physics, Northeastern University, 
Boston, MA 02115, USA}
\address{$^{2}$European Synchrotron Radiation Facility, 
BP220 Grenoble, 
France}
\date{\today}
\maketitle

\begin{abstract}
Recent x-ray Compton scattering experiments in ice 
have provided useful information about
the quantum nature of the interaction between H$_2$O monomers.
The hydrogen bond is characterized by a certain amount of charge
transfer which could be determined in a Compton experiment.
We use ab-initio simulations to investigate the
hydrogen bond in H$_2$O  structures by calculating the Compton 
profile and related quantities in three different systems, 
namely the water dimer, a
cluster containing 12 water molecules and the ice crystal.
We show how to extract estimates of the charge transfer
from the Compton profiles.
\end{abstract}


\vskip 1cm

\section{Introduction}
The hydrogen bond is one of the least well-understood components in the 
energy decomposition that is used to predict the folding of 
biological complexes such as proteins.
Its importance stems from its directionality and 
modest bonding 
energies midway between strong covalent and weak Van der Waals bonds.
For this reason the inter-molecular 
interaction is difficult to characterize. 
Recently, X-ray Compton scattering measurements in ice
revealed subtle oscillations 
in the profile anisotropy,
which can be interpreted as reflecting various
quantum mechanical aspects of the hydrogen bond 
\cite{prl_ice,isaacs2,shukla02,romero,ragot}.
In simple descriptions of the hydrogen bond 
\cite{ghanty,coulson,bratoz},
the highest occupied molecular orbital
(HOMO) of a proton acceptor momomer can be viewed
as a lone pair $p$ state of the oxygen, while
the lowest unoccupied molecular orbital (LUMO)
of the donor neighbor molecule 
is the  antibonding OH orbital. 
Therefore, the problem can be reduced to an effective  
2 by 2 eigenvalue calculation
where the HOMO-LUMO mixing can in principle be measured 
by an angle $\theta$.
In this effective model the amount of charge transferred can 
be defined as
\begin{equation}
\Delta Q 
=\int 
|\cos \theta <{\bf r}~ |~0 > +
  \sin \theta <{\bf r}~|~ 1 > - 
 <{\bf r}~|~0>|^2~d^3 
{\bf r}
\end{equation}
where $<{\bf r}~|~0 >$ is the wavefunction of the lone pair
and $<{\bf r}~|~1 >$ is the wavefunction of the LUMO
at the donor neighbor molecule. When the mixing 
is small, one obtains
$\Delta Q \approx \sin^2\theta$. 
Besides, in this limit,
$\sin\theta$ turns to be approximately
proportional 
to the energy gain due to the hydrogen 
bond. 

Compton scattering gives us information on
the (ground state) electronic momentum distribution
of the system under study. Of particular interest
is the width $ \Delta p $ of the momentum density over
which the valence momentum density falls to zero around
a cutoff momentum (Fermi momentum in a metal).
Since the LUMO contains higher momentum components
than the HOMO, the variation of the width $ \Delta p$ 
is clearly related to the mixing $\theta$.
Moreover, Friedel and
 Peter\cite{friedel}
provided a useful
discussion on the impact in momentum density of 
the mixing of occupied with virtual orbitals.
Their arguments can be extended to the present case 
in order to
estimate the charge transfer mixing angle.
We limit our discussion here to the more general concept of 
charge transfer without invoking the term covalency \cite{ghanty}.
As a matter of fact ionic compounds are also 
characterized by a charge transfer but
they are certainly not described as covalent systems.

This paper is organized as follows. In Sec. II
we present the computational method and recall
the definitions of the quantities studied.
Sect. III is devoted to estimate the charge transfer 
and to describe its properties in position space.
Finally, Sec IV summarizes our main conclusions. 

\section{Method}
We start with the one-particle density matrix, 
$\hat{\rho}({\bf r},{\bf r}^{\prime})$,
which is defined in terms of
the normalized N-particle wavefunction, $\Psi$, as
\begin{equation}
\hat{\rho}({\bf r},{\bf r}^{\prime})
=N\int d\xi 
~\Psi^*({\bf r},\xi)\Psi({\bf r}^{\prime},\xi)~. 
\nonumber
\end{equation}
where the integral extends over the coordinates 
of all other particles.
If the many body wavefunction is represented by a single 
determinant, which is true in the case of Hartree Fock 
then the density matrix is idempotent $(\hat{\rho}=\hat{\rho}^2)$ 
and reduces to a summation over the occupied spin-dependent 
orbitals $\psi_{i}$, i.e.
\begin{equation}
\hat{\rho}({\bf r},{\bf r}^\prime) = 
 \sum_{i=1}^{N} \psi_i({\bf r})\psi^*_i({\bf r}^\prime)\,.
\end{equation}
The Electron Momentum Density (EMD) is 
defined as the momentum transform of the density matrix, 
\begin{equation}
\rho({\bf p})= \frac{1}{8 \pi^3}
\int \int d^3{\bf r} d^3{\bf r}^\prime
\hat{\rho}({\bf r},{\bf r}^\prime) 
\exp(-i {\bf p} \cdot ({\bf r} - {\bf r}^\prime))~.
\end{equation}
Note that, in general, the EMD involves off-diagonal elements 
of the real space density matrix.
In connection with the Compton spectra, it should 
be noted that the Compton profile (CP), $J(p_z)$, represents 
the double integral of the ground-state EMD
$\rho ({\it\mbox{\boldmath$p$}})$:
\begin{equation}
\label{eq_cp}
J(p_z) = \int\!\!\!\int\rho ({\bf p})dp_x dp_y,
\end{equation}
where $p_z$ lies along the scattering vector of the x-rays. 

\begin{figure}[htb]
\unitlength=1cm
\begin{center}
\begin{picture}(7,8.3)
\put(-2.0,-1.0){\epsfysize=9cm
\epsffile{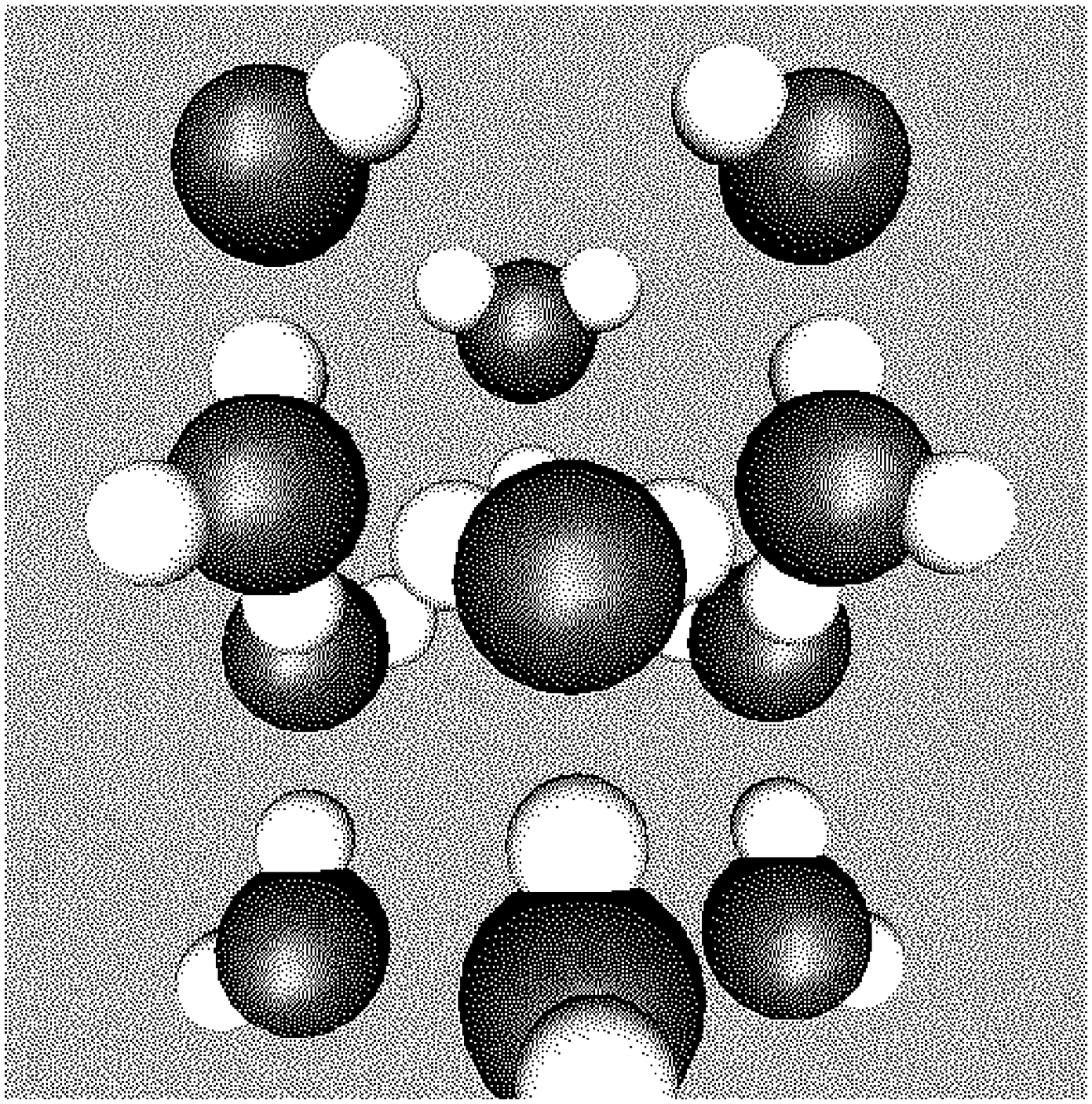}}
\end{picture}
\end{center}
\caption{Water cluster containing 12 water molecules.} 
\label{cluster}
\end{figure}

The program used in the present work was CRYSTAL98 \cite{crystal98},
which is especially appropriate for 
first-principles calculations of the density matrix 
$\hat{\rho}({\bf r},{\bf r}^{\prime})$
in extended (periodic) and in molecular systems.
For the ice crystal we have used the experimental atomic distances.
We have also considered the dimer and cluster 
containing 12 water molecules 
which are shown in Fig. \ref{cluster}. 
The O-O distances and the orientations of the atoms in 
the cluster and in the dimer are equal
to the corresponding ones in the crystal.  
The CRYSTAL98 program employs linear combination 
of atomic orbitals for
solving the Hartree-Fock equation.
One can show that the 
wavefunction involved in the 
hydrogen bond are described reasonably 
well already at the level of
the Hartree-Fock approximation \cite{vanq,ojamae}. 
The occupied orbitals used to determine 
the momentum density and 
the Compton profile are calculated using 
the Restricted Hartree-Fock 
(RHF) scheme with the 6-31G$^{**}$ 
basis set \cite{ojamae}.

\section{Results}
\begin{figure}
 \epsffile{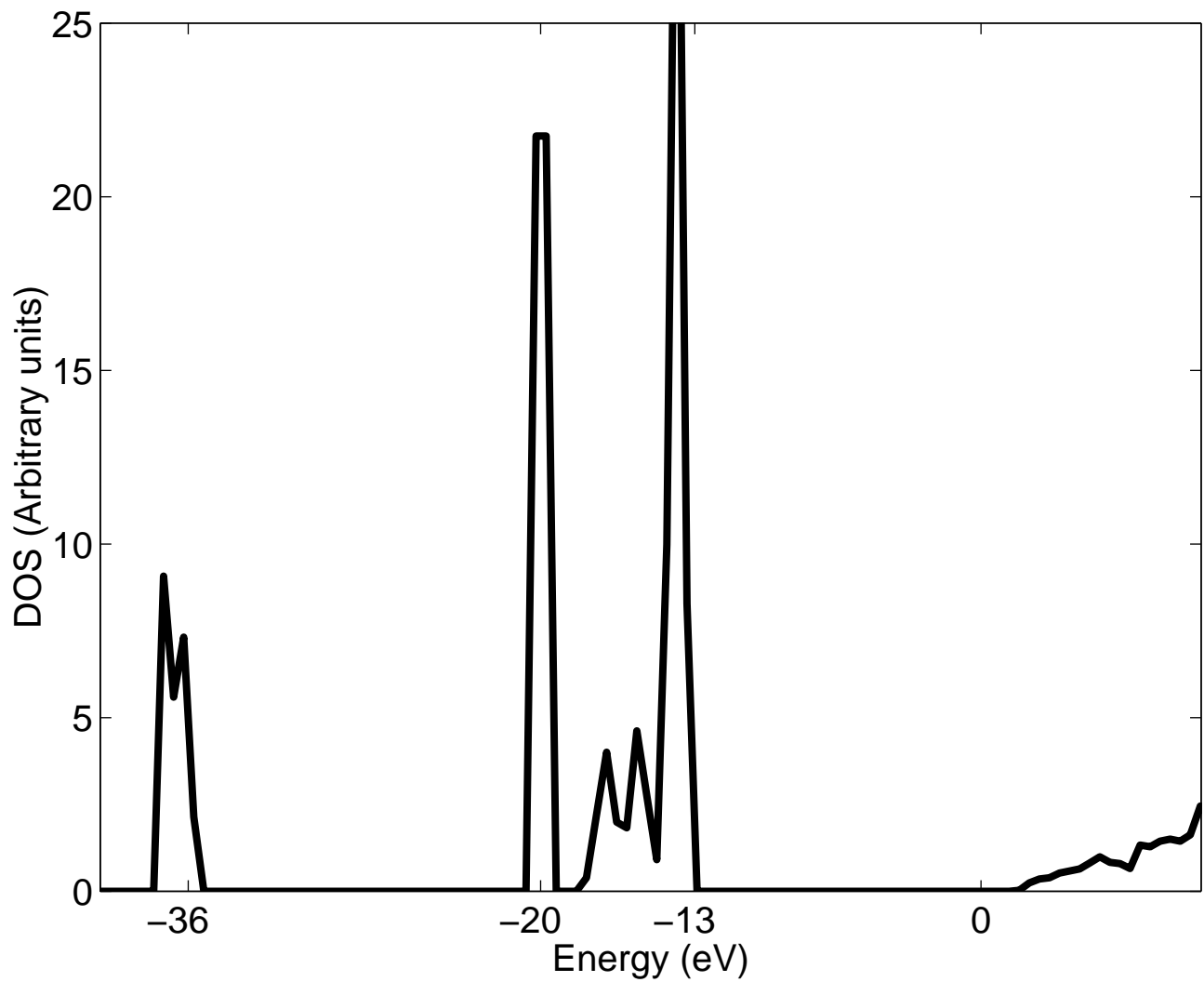}
  \caption{DOS of the ice crystal according to the RHF calculation.} 
  \label{dos}
\end{figure}

\begin{figure}
 \epsffile{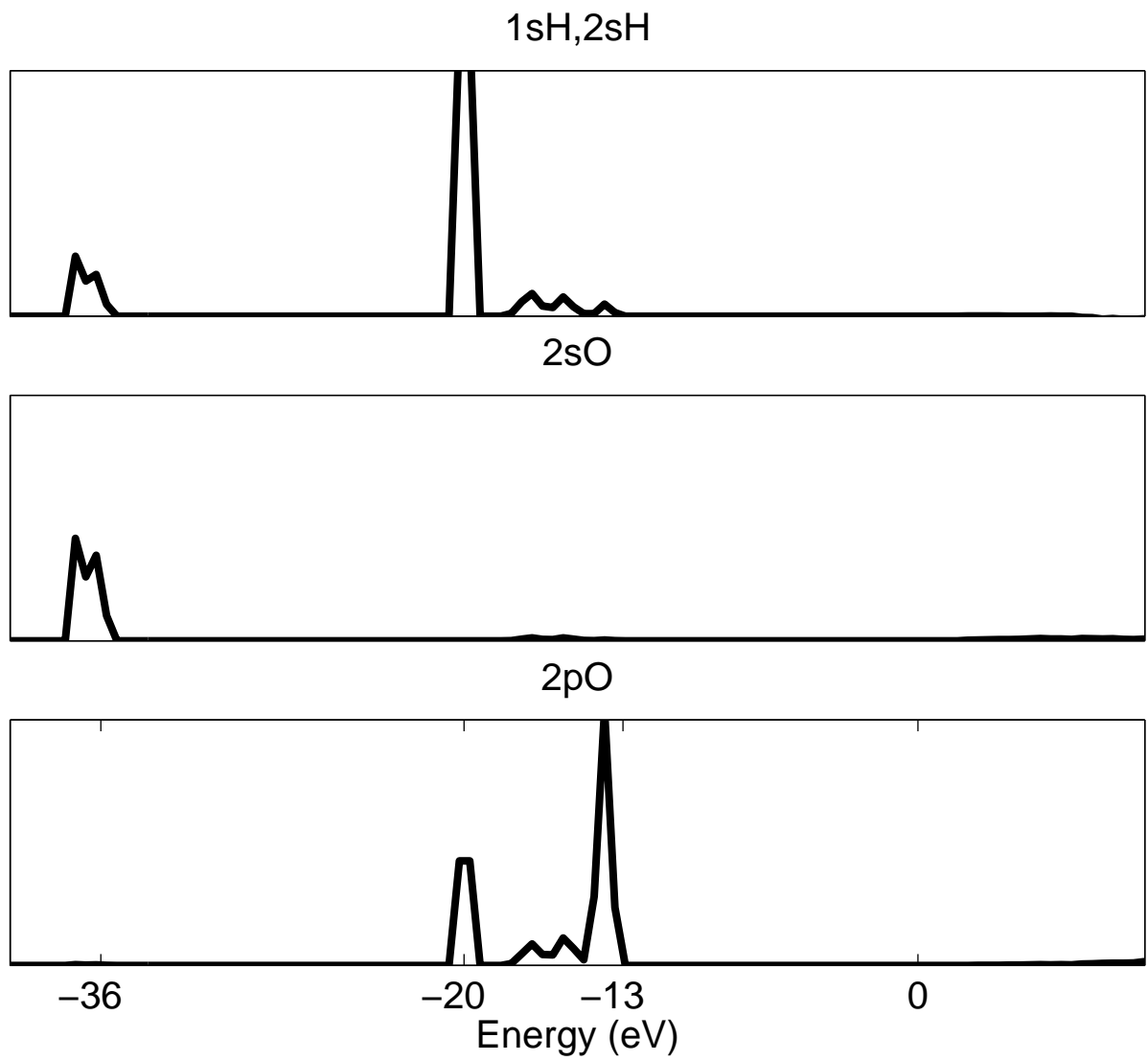}
  \caption{Projected DOS per orbital of the ice crystal according to the RHF calculation.} 
  \label{pdos}
\end{figure}

The density of state (DOS) of the ice crystal is 
shown in Fig.\ref{dos}. The band gap is about 15 eV. 
In the occupied bands, one can notice two high
peaks near -20 eV and near -13 eV. 
For comparison,  in the dimer the calculated
RHF energies of the 8 valence occupied states are 
-37 eV, -36 eV, -20 eV, -19 eV, -16 eV, -15 eV, -14 eV and -13 eV. 
The electronic density of these dimer states
has been visualized by Chaplin \cite{chaplin}.
One can see that the densities of the orbitals 
corresponding to -19 eV and to -15 eV
show overlap across the hydrogen bond.
Interestingly, 
the two peaks of the crystal DOS correspond to dimer orbitals
orthogonal to the hydrogen bond:
$-20$ eV corresponds to a state strongly localized on
the acceptor molecule while $-13$ eV
corresponds to a state localized on the donor
water molecule.
Clearly the dimer has some 
very important things to tell us about the nature of 
the hydrogen bond.
The projected DOS shown in Fig.\ref{pdos}
reveals that the peaks near $-20$ eV is a combination
of 2pO and (1sH,2sH) while the peak near $-13$ eV
is almost a pure 2pO.
Moreover, one can notice that 
the highest occupied bands are mostly a combination
of 2pO and (1sH,2sH) states 
with a very small 2sO contribution.
This picture is therefore consistent
with the charge transfer theories \cite{coulson,bratoz} 
explaining the hydrogen bond by
the mixing of the oxygen lone pair 
with an antibonding OH orbital 
centered on a neighbor molecule.

In order to extract molecular energies 
and other properties, it can be useful 
to build a  density matrix as a superposition 
of density matrices of isolated molecules 
arranged in the same geometry as 
in the crystal or the cluster. 
The keyword MOLSPLIT (in the CRYSTAL 98 input) 
performs an expansion of the lattice, 
in such a way that the molecules of the system 
are at an infinite distance 
from each other to avoid inter-molecular interactions.  
Our calculations indicate that the gain in energy 
per hydrogen bond 
in the crystal is $-0.5$ eV.
This energy gain is strengthened by
a factor of two as one goes up in the cluster size.

\begin{figure}
 \epsffile{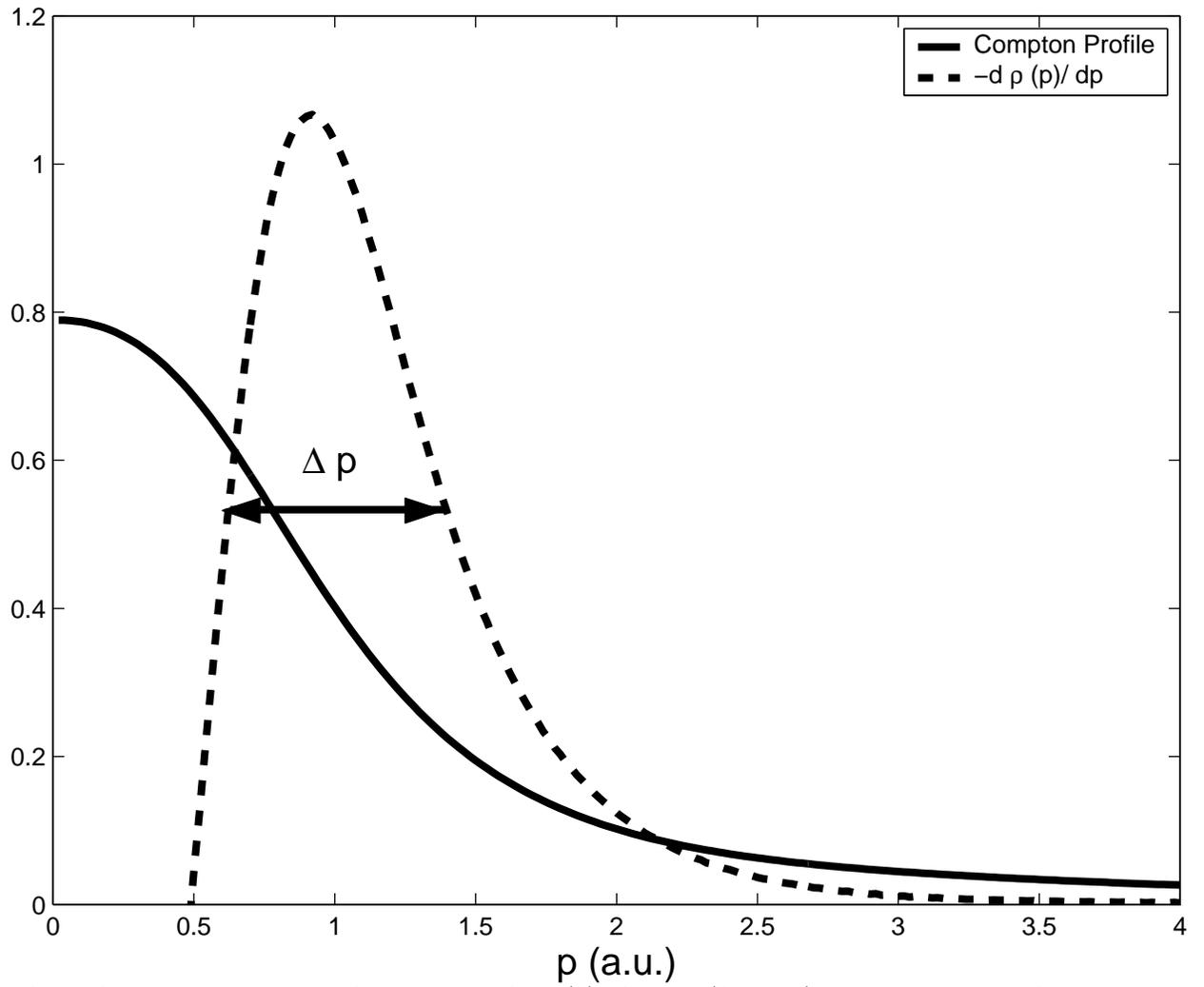}
  \caption{Spherically averaged Compton profile $J(p)$ of water (full line)
  together with the function 
$-d n(p)/ dp$ (dashed line) used to define
  the quantity $\Delta p$.} 
  \label{CPSA}
\end{figure}

Our aim in this work is to probe theoretically the effects
of the hydrogen bond in water by evaluating the changes
it provokes in the electronic momentum distribution which
can be probed using an experimentally accessible quantity, 
the Compton profile. For this we use the spherical average 
of the momentum density $n(p)$ which can be obtained from the
isotropic, or spherically averaged Compton profile $J(p)$
by the formula
\begin{equation}
n(p)=\frac{1}{p} \frac{dJ(p)}{dp}~.
\end{equation} 
The spherical averages $J(p)$ and $n(p)$ 
are thus used for the evaluation of  a quantity $\Delta p$,
defined as the full-width at half maximum of 
the  broad maximum exhibited by the function 
$-d n(p)/ dp$ (see Fig. \ref{CPSA}). 
This maximum corresponds to the typical valence 
cutoff momentum (the Fermi momentum in the case of a metal).
In the limit of a small mixing angle, 
formulae similar to those 
given by Friedel and
 Peter\cite{friedel} 
indicate that $\sin \theta$ is
proportional to the variation of the momentum density 
smearing
 width.
 Moreover, by assuming a linear scaling between
the momentum width and the energy separation
of the effective 2 by 2 eigenvalue problem, 
we obtain
\begin{equation}
\sin \theta\approx \frac{1}{\sqrt{2}}
\frac{\Delta p- \Delta p_0}{\Delta p_0}~,
\label{theta}
\end{equation}
where $\Delta p_0$ corresponds 
to the single water molecule and $\Delta p$
to hydrogen bonded water 
structure being probed.

\begin{figure}
 \epsffile{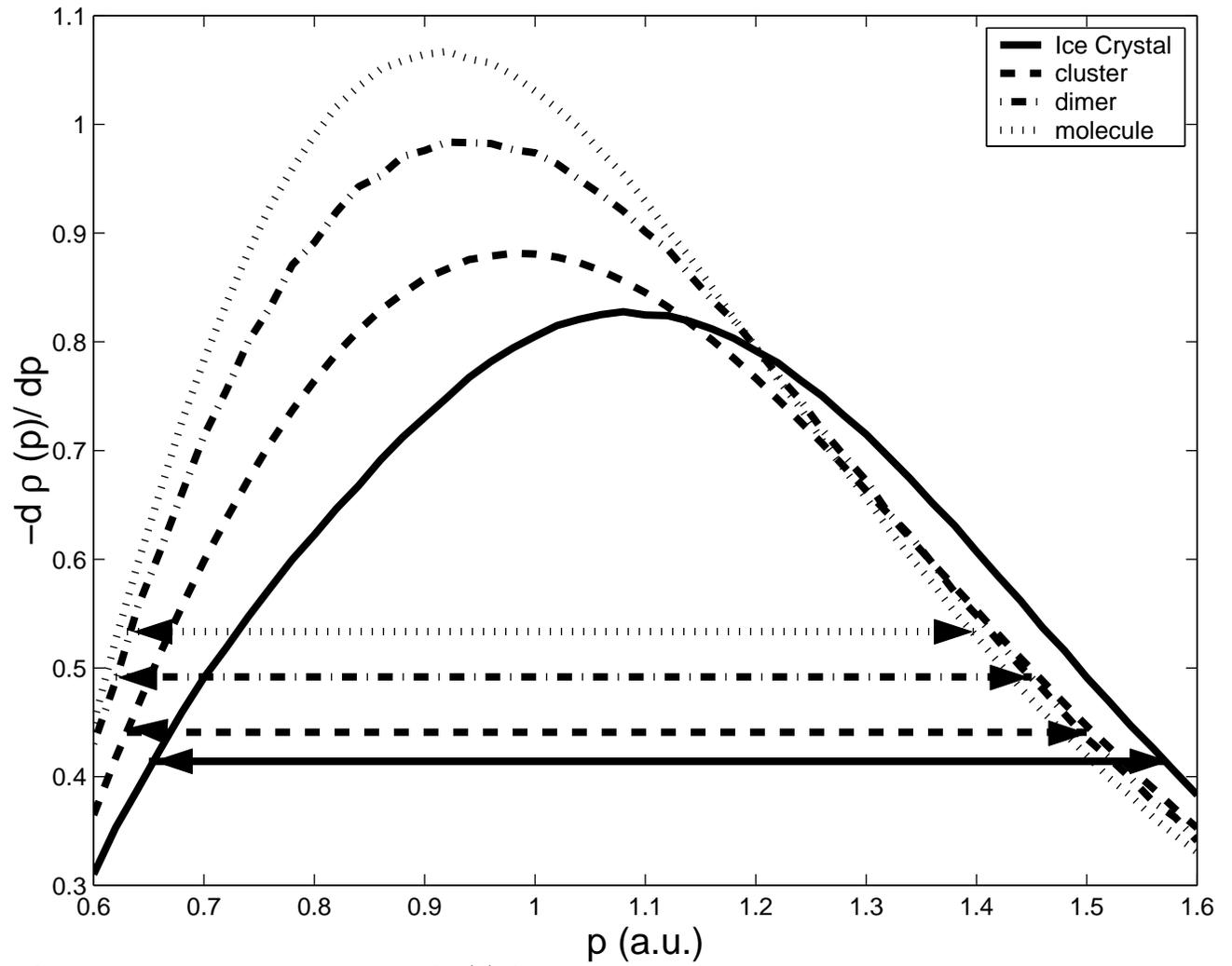}
  \caption{Negative radial derivative of $n(p)$
   for the water molecule, the dimer, the
   cluster containing 12 water molecules and the ice crystal.} 
  \label{DN}
\end{figure}

The radial derivative of $-d n(p)/ dp$ is shown in   
Fig. \ref{DN} for the water molecule, the dimer, the
cluster containing 12 water molecules and the ice crystal.
The  increase of $\Delta p$ as a function of the cluster 
size is striking. 
The $\sin \theta$ given by Eq. (\ref{theta})
it is strengthened by
a factor of two 
as one goes from the dimer to the crystal.
Thus, these estimates yield 
values consistent with the trends 
observed 
in the total energy
calculations.
The amount of charge transferred is about $0.5 \%$ 
of an electron in the dimer
, $1 \%$ in the cluster
and $2 \%$ in the crystal.
Interestingly these values are in reasonable 
agreement with 
other estimates of the charge transfer given
in previous quantum mechanical calculations 
\cite{coulson,bratoz,barnett}. 
Finally, in Fig. \ref{DN} 
we also notice a shift of the
maximum of the function $-d n(p)/ dp$ as a 
function of the 
cluster size, which can explain
an observed increase of the kinetic 
energy. 

\begin{figure}
 \epsffile{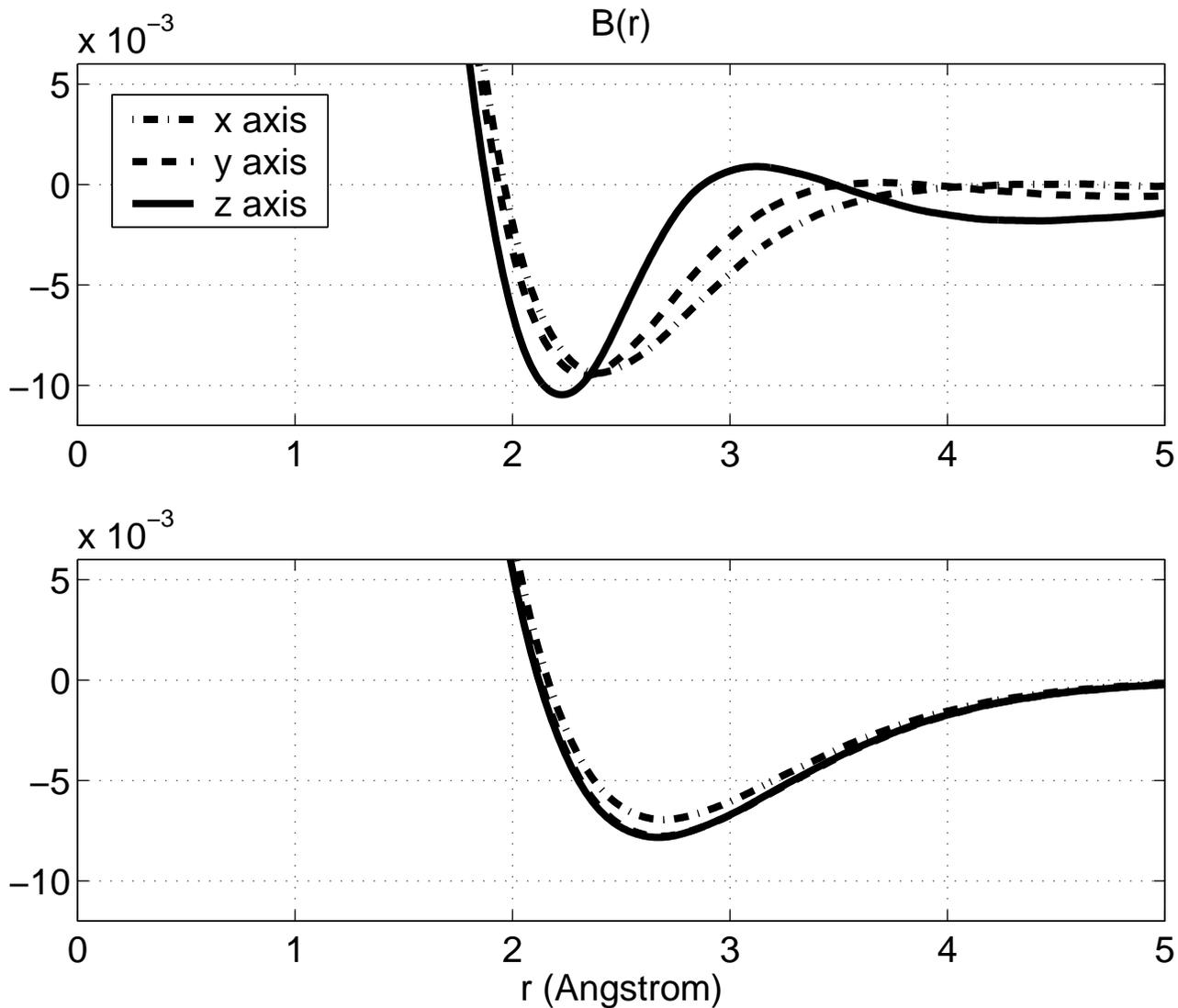}
 \caption{Top: theoretical $B({\bf r})$ for the Ice crystal along $x$ 
 (dashed-dot line),
 $y$ (dashed line) and $z$ (full line) directions obtained from the Compton
 profiles calculated by CRYSTAL98 \protect \cite{crystal98}.
 Bottom: corresponding lines for the isolated molecules.
 The lines are normalized so that $B({\bf 0})=1$.}
 \label{BR}
\end{figure}

\begin{figure}
 \epsffile{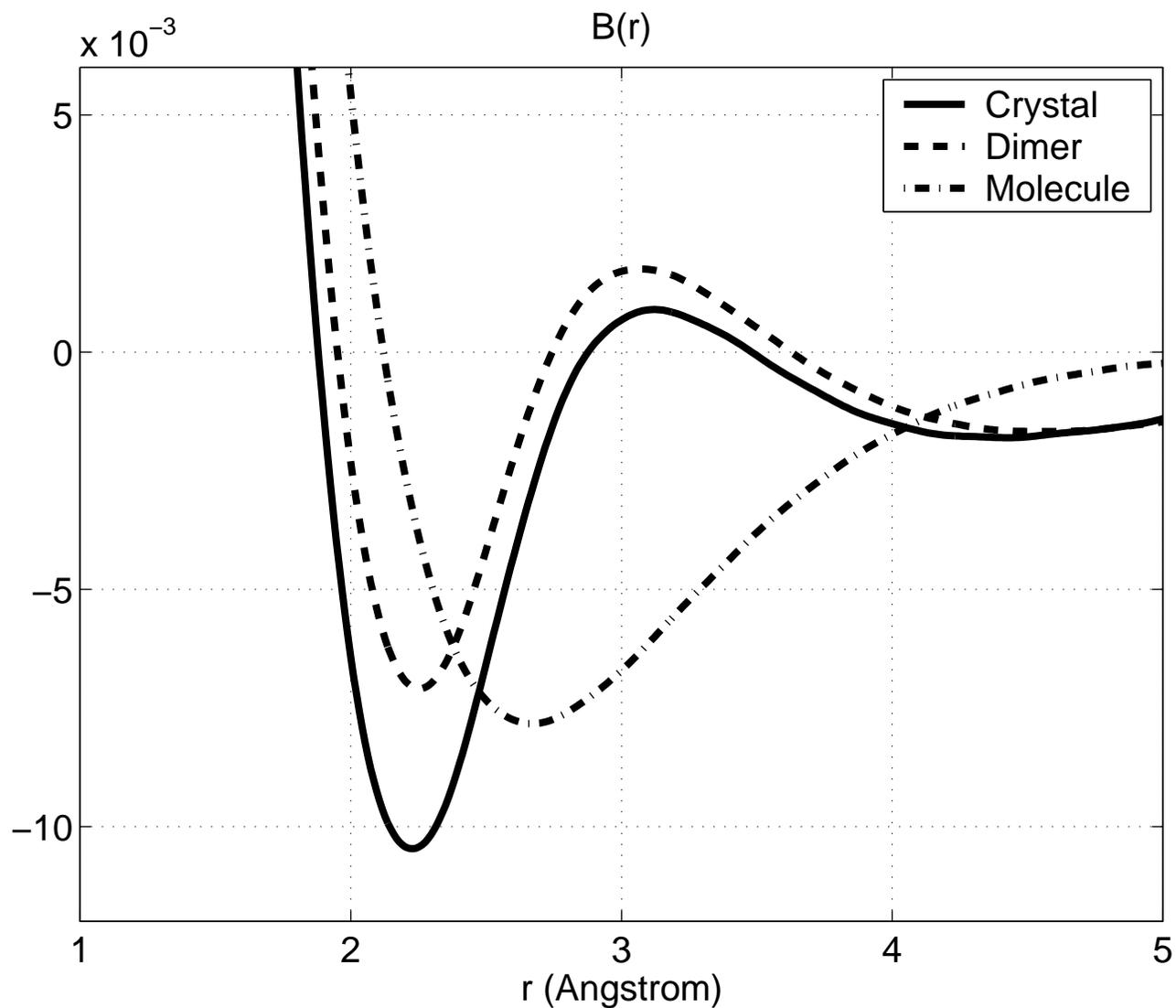}
 \caption{Top: theoretical $B({\bf r})$ for the Ice crystal (full line)
 for the dimer (dashed line) and the water molecule (dashed-dot line)
 in the direction of the strongest hydrogen bond.
 The lines are normalized so that $B({\bf 0})=1$.}
 \label{BZ}
\end{figure}

In order to show how Compton scattering data can assist
in the interpretation of the bonding properties in position
space, one can define a function $B({\bf r})$ which is the 
Fourier transform of the momentum density 
\cite{pattison1,pattison2},
\begin{equation}
B({\bf r})=\int \rho({\bf p}) 
\exp(-i{\bf p}\cdot{\bf r})~d^3 {\bf p}~.
\end{equation}
From the convolution theorem,
$B({\bf r})$ is just the autocorrelation of the one-electron 
wave functions. Moreover $B({\bf r})$ in a given direction
can be calculated as the Fourier transform of the Compton profile
in this direction.
The orbital contribution of the 1s core electron of oxygen
vanishes beyond 1.5 \AA\  so that the $B({\bf r})$ is dominated
by valence electrons.
By analyzing $B({\bf r})$ in the direction $z$ parallel to 
the hydrogen bond,
we observe a negative local minima near 2 \AA , a positive 
local maximum
near 3 \AA\  and finally a negative local minima at higher 
distances as shown in Fig.\ref{BR}. 
The corresponding lines along $x$ and $y$, also plotted in
in Fig.\ref{BR}, differ substantially
from the $z$ direction.
Both $x$ and $y$ curves are quite similar and present only the
local negative minima near 2 \AA, which is broader 
toward the large distances, 
and then their amplitudes go to zero at larger distances.
The total autocorrelation function $B({\bf r})$ can be also
interpreted in relation to the hydrogen bond with a comparison 
to isolated aligned molecules 
(obtained by MOLSPLIT). The isolated water molecules 
curves are shown in the bottom part of Fig.\ref{BR}
along $x$, $y$ and $z$. 
One can observe that none of these curves presents
the oscillations 
that characterizes the $z$ direction in the crystal. 
The autocorrelation function $B({\bf r})$ along 
the direction
of the strongest hydrogen bond is shown in Fig. \ref{BZ} for
the crystal and the dimer. Both curves deviate strongly from
the monomer curve. They also present some clear differences
among them: the crystal possesses a deeper minimum 
and a somewhat lower maximum at 3 \AA. 
However, beyond 4 \AA\  the two
curves cannot anymore be distinguished indicating that
the charge transfer occurs mostly within this distance 
range.

\section{Conclusion}
In conclusion, this study confirms that
the Compton profile is a useful tool for the
investigation of the quantum nature of hydrogen bond
in ice \cite{prl_ice} and probably in other
compounds such as 
urea \cite{urea},
RNA and DNA \cite{weyrich}. 
Our results show a unique sensitivity
of Compton scattering to the momentum width $\Delta p$.
In this work an attempt has been done to give an
estimate of the HOMO-LUMO mixing 
from Compton profiles.
The present method can be also used for amorphous 
materials
 where Compton profiles anisotropies
are not available.
Moreover, we have used the directional Compton profiles 
to study the bonding properties in the position
space by using the autocorrelation functions.
We have determined that the charge
transfer occurs mostly within a range of about 4 \AA\ .
The present work also shows that x-ray inelastic scattering 
experiments can shed light on the important role
in short hydrogen bonds played by
quantum mechanical charge transfer between nearest
molecules. 
These phenomena cannot
be described by the standard effective classical potentials
used for instance in some water clusters 
simulations \cite{liu}.

We wish to acknowledge enriching 
discussions with P.M. Platzman, D.R. Hamann,
E.D. Isaacs, W. Weyrich, N. Marzari and M. Boero.
We also thank S. Ragot for sending us ref. \cite{ragot},
Y. Zhukovski and S. Piskunovs for helping
us to visualize the projected DOS and
A. Gull\`a for a careful reading of the manuscript.
This work is supported by the US Department of Energy under
Contract No. W-31-109-ENG-38, and benefited from the 
allocation of supercomputer time at the Northeastern 
University Advanced Scientific Computation Center (NU-ASCC).

\newpage

\end{document}